\documentclass[fleqn,10pt]{arxiv}

\newcommand{\completeTest}{{\sc CompleteTest}}
\newcommand{\uppaal}{{\sc Uppaal}}
\usepackage{graphicx}
\usepackage{caption}
\usepackage{subcaption}

\begin{document}
\title{Enablers and Impediments for Collaborative Research in Software Testing: An Empirical Exploration}

\author[1]{
Eduard Paul Enoiu, Adnan \v{C}au\v{s}evi\'{c}}

 \address[1]{Software Testing Laboratory, M\"alardalen University, Sweden}
 
 \ead{eduard.paul.enoiu@mdh.se, adnan.causevic@mdh.se}

\maketitle
\begin{abstract}
When it comes to industrial organizations, current collaboration efforts in software engineering research are very often kept in-house, depriving these organizations off the skills necessary to build independent collaborative research. The current trend, towards empirical software engineering research, requires certain standards to be established which would guide these collaborative efforts in creating a strong partnership that promotes independent, evidence-based, software engineering research.
This paper examines key enabling factors for an efficient and effective industry-academia collaboration in the software testing domain. A major finding of the research was that while technology is a strong enabler to better collaboration, it must be complemented with industrial openness to disclose research results and the use of a dedicated tooling platform. We use as an example an automated test generation approach that has been developed in the last two years collaboratively with Bombardier Transportation AB in Sweden.

\end{abstract}

\section{Introduction}
Technological, organizational and economic factors influence profoundly the quality of software engineering research worldwide. Software testing is one of the biggest research direction in software engineering. Wong et al. \cite{glass1994assessment} indicated that for 37\% of the top scholars in software engineering, their research focus includes {\it software testing}.
In this area, but not restricted to it, there is presently a mismatch between an industrial problem and the technical capacity of companies to generate and use new and existing knowledge for overcoming this burden.

Software testing research provides a case for technologies, methods, and knowledge invoking changes in companies. Two major groups of entities have been involved in the industrial software testing research: universities and companies. Both entities have made important contributions to the progress of this research area. While there have been a number of studies looking at {\it collaborative research} in different domains \cite{lee2000sustainability,dooley2007university,lee1996technology,perkmann2007university}, there are few studies which have looked at enabling factors for a success collaborative research in software testing. Sigrid Eldh \cite{eldh2013some} has investigated research collaborations and how to avoid problems by considering solution approaches, communication focus, unpleasant results, mismatching interpretations, business aspects and an evaluation criteria. 

In what follows, one collaboration of Bombardier Transportation AB, a leading, large-scale company focusing on development and manufacturing of trains and railway equipment is described. The results of this collaboration are based on industrial-scale problems when using an {\it automated testing} approach as part of the testing process in the involved organization. This case provides input to our observations and offers a longitudinal opportunity for exploring and fostering interest in automated test generation for industrial software systems. In addition, it indicates some key {\it enabling factors} as well as potential {\it impediments} of a successful industry-academia collaboration in software testing research.

\section{A Case Study at Bombardier \\ Transportation}
\begin{figure}
\centering
\includegraphics[width=\textwidth]{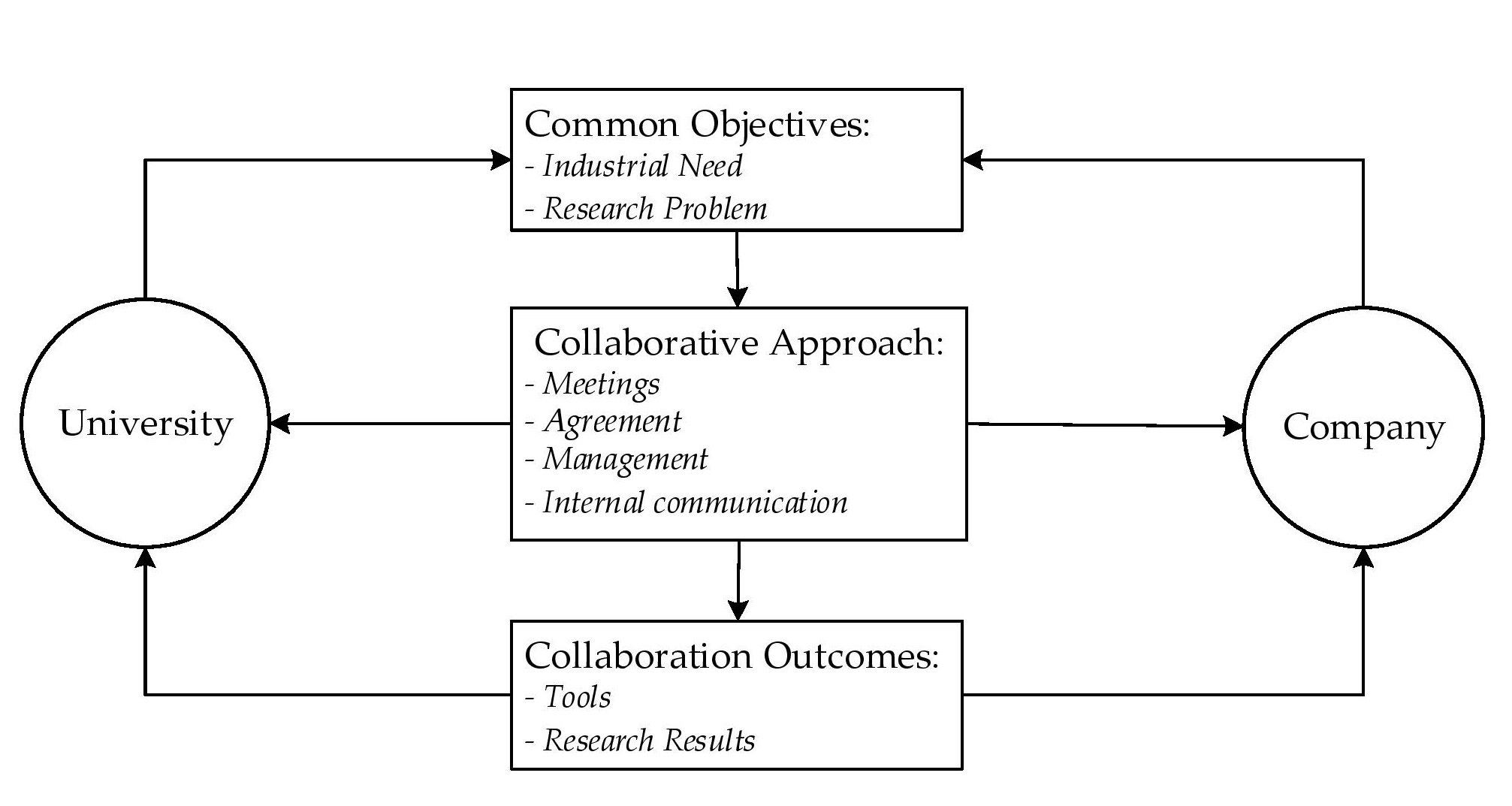}
\caption{Model of Collaborative Research for the Case Study}
\label{fig:model}
\end{figure} 

In software engineering, test engineers are required to validate the software against their specifications as well as to show that tests exercise, or cover, the structure of the software. Use of automated test generation techniques has been proposed by several researchers \cite{ali2010systematic}. The limited
application to real-world industrial projects, however, impacts the transfer of test generation technologies and hence its impeding the collaboration efforts between industry and academia. Thus, there is a need to validate these
approaches against relevant industrial systems such that more knowledge is built on how to efficiently use them in practice.

We started a research collaboration with Bombardier Transportation AB, a train manufacturing company that produces both hardware and software. At Bombardier, the test specification, execution and reporting are performed by a test engineer. Before starting the collaboration test engineers were manually developing tests. During this collaboration, we showed how to efficiently and automatically generate tests for a train control software system \cite{enoiu2013model,enoiu2013using,enoiu2013mos}. We developed a toolbox, named {\completeTest}\footnote{{\completeTest} - available at www.completetest.org.}, suitable for industrial applications. The main goal for the design of the toolbox was to meet the exact needs of an industrial end user. Although there is a possibility for fine tuning the configuration parameters of the underlying techniques, most of them are set to default values, making the technique immediately ready for usage upon startup.

Perkman \cite{perkmann2008engaging} is distinguishing between three types of collaborative research: opportunity driven, commercialization-driven and research-driven.
In 2012, a collaborative research-driven collaboration was established between Bombardier Transportation AB and M\"alardalen University both located in V\"aster{\aa}s, Sweden. As shown in Figure \ref{fig:model} this cooperation is driven by our common research opportunities and a long-term relationship. The vision of this collaboration is to improve the state of the practice in automated test generation and evaluation through design, implementation and conduct of relevant research that could be translated into software testing policy and practice. A major emphasis was made on using available research in the area of {\it model-based testing}. 

\begin{figure}[tbph]
\centering
\includegraphics[scale=0.66]{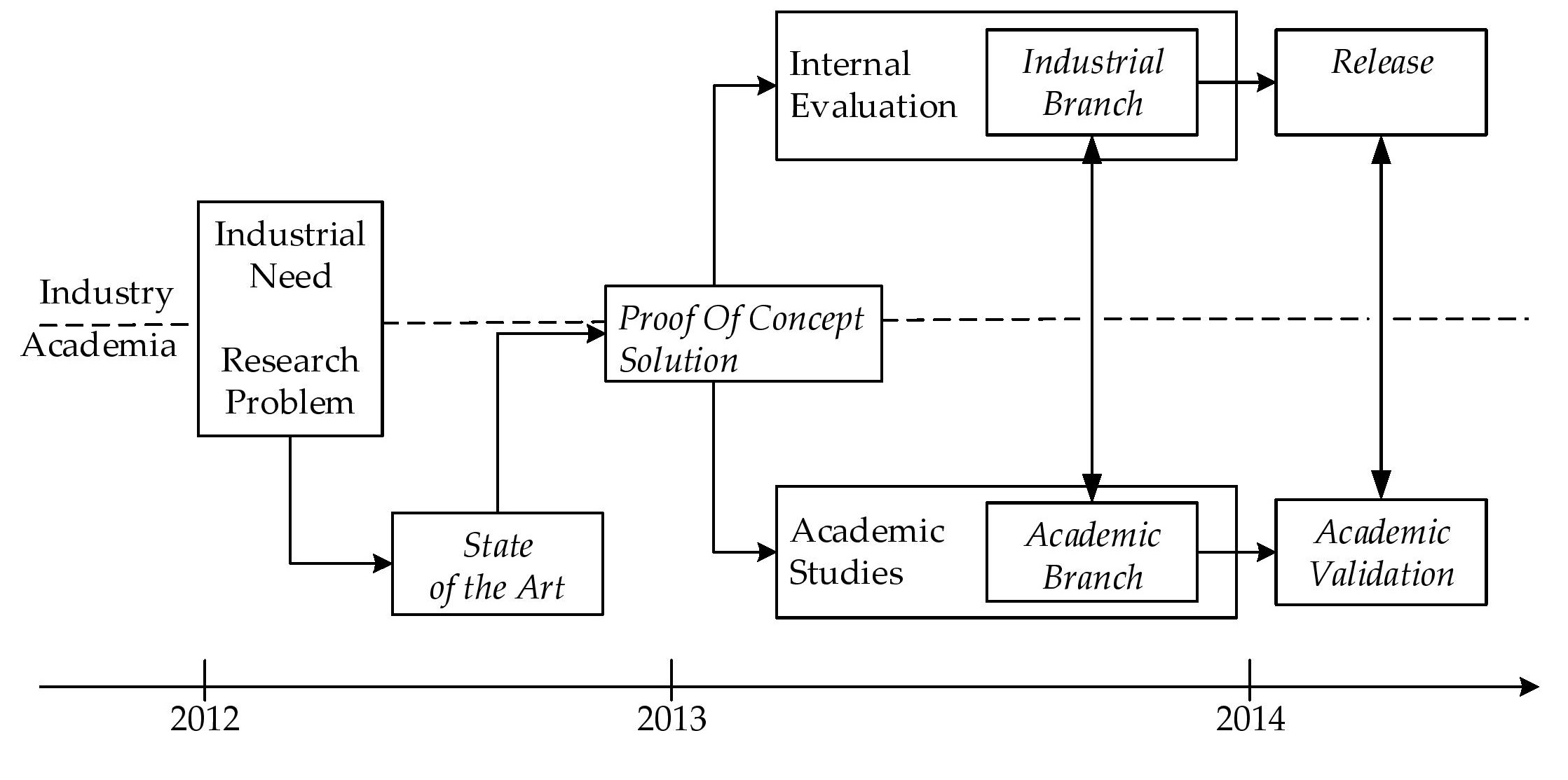}
\caption{Case Study Roadmap}
\label{fig:timeline}
\end{figure}

Collaborative research often occurs with academia and universities. Although collaboration starts in many different forms and motives, a generalizable model underpins research collaborations in the sense used in this case study. As shown in Figure \ref{fig:model} the partnerships was build upon {\it common objectives}. Both partners were keen to demonstrate the industrial efficacy of the new and uncertain automated test generation technology. The collaborative approach demonstrates that the university and the company can together obtain tools and applied research results which they could not achieve independently. Secondly, it can facilitate the transfer of research know-how both from collaboration outcomes but more importantly during the collaborative approach through internal communication, meetings and mutual agreements.

This scientific research was conducted within the ATAC project (Advanced Test Automation for Complex and Highly Configurable Software-intensive Systems) started in 2012 by 15 European partners and then continued through the ITS-EASY research school. Due to the cross-organizational nature of this effort, resources were pulled together from both participating organizations, enabling a PhD student to dedicate 40\% of his academic time to industrial project development. The project aim was to develop, enhance, and deploy high performance methods and tools for automated quality assurance of large and distributed software-intensive systems.

\begin{table}[ht]
\renewcommand{\arraystretch}{1.0}
\caption{Example of an Action Plan for Piloting the {\completeTest} Industrial Branch}
\label{tab:action}
\centering
\begin{tabular}{p{0.06\linewidth} p{0.5\linewidth} p{0.2\linewidth}}
    \hline
Step & Description & Responsibility\\
    \hline
    \\
1 & Investigate the proof of concept solution and map to requirements to identify gaps. & U  \\
    \hline
    \\
2 & Pilot the use on selected software to understand the consequences and required changes for developers and test engineers. & U \& C  \\
    \hline
    \\
3 & Develop industrial branch and pilot the solution in a lab environment. & U \& C  \\
    \hline
    \\
4 & Plan for the use of the new solution and implement them in a pilot project. & U \& C  \\
    \hline
    \\
5 & Introduce the new methods and tool throughout the organization (i.e., roll out for each new project). & C  \\  
    \hline

\end{tabular}
\end{table}

The work performed in this case study is based on theories of formal verification, model checking and model-based test generation. Through this case study the {\completeTest} tool and these methods have been made accessible to engineers and developers in Bombardier Transportation AB. As shown in Figure \ref{fig:timeline} we started in the beginning of 2012 by targeting a {\it research problem} and an {\it industrial need}. We performed a state of the art study \cite{atacsota} on this research problem and found that by employing a novel solution that facilitates model-based test suite generation for control systems we can improve the state of the practice significantly. We investigated in 2013 a {\it proof of concept} solution and devised a plan seen in Table \ref{tab:action}\footnote{In Table \ref{tab:action} U represents University and C the Company} for ensuring good communication and sharing of information between both partners. In the same time a couple of academic studies were performed during 2013 with results published in peer-reviewed workshops and conferences \cite{enoiu2013model,enoiu2013using,enoiu2013mos}. Based on these results we devised an academic branch of the solution that contains more features compared to the industrial branch (e.g., mutation testing, test optimization). From the second semester of 2013 to the beginning of 2014 test engineers from the company collaborated intensively with us to transfer the knowledge directly from the academic studies to the existing company practice. The industrial release features a completely new process for automated test generation including {\it tool support}. Engineers observed that in the new process the potential for saving testing time is enormous when considering the amount of manual testing that has been done to satisfy safety standards for development of safety software. Process improvement measurements are subject to further academic studies during 2014 including both testing efficiency and effectiveness. 

In 2014 in order to validate the result, extensive empirical studies of the method, where the tool is applied on real-world industrial programs, are currently under development. Through these studies, we continue to validate the proposed approach against industrial systems, thus building more knowledge on how to efficiently test them in practice.

\section{Enabling Factors}
\begin{table}
\renewcommand{\arraystretch}{1.0}
\caption{Observed Enablers of Collaboration Research.}
\label{tab:study}
\centering
\begin{tabular}{p{0.01\linewidth} p{0.85\linewidth}}
    \hline
    \\
1 &  Open sharing of information for research purposes. High perceived "value" in sharing information for research purposes.    \\
2 & Use of a dedicated tooling platform. Development of both academic and industrial branches such that different goals can be achieved. \\
3 & Ensure weekly collaboration through dedicated progress meetings.   \\
4 &  Create the research culture and promote the benefits internally.\\
    \\
    \hline
\end{tabular}
\end{table}
This paper looks at the enabling factors in collaborative research between industry and academia based on our experiences during the {\completeTest} case study. Table \ref{tab:study} depicts the observed benefits of our specific research collaboration. In general, it is rather difficult to tell whether, in this particular case, success was achieved by the actual technology or the applied transfer process. We believe, that both the method and the technology transfer process are influencing the success of our case study and depends upon the commitment of those involved (i.e., both academic and industrial members). Nevertheless, the company employed the collaboration outcomes into their practice not as easily as expected and the endeavor takes substantial time and effort.

\subsection{Open Sharing of Information}
In software engineering research, academics are very depended on industrial projects which can be sometimes at odds with modern open science practices. Czarnitzki et al. \cite{czarnitzki2014delay} investigated the relationship between industry sponsorship and restrictions on disclosure and indicated that industry sponsorship is a {\it real threat} to public disclosure of academic research.

\begin{figure}[tbph]
        \centering
        \begin{subfigure}[b]{\textwidth}
                \includegraphics[width=\textwidth]{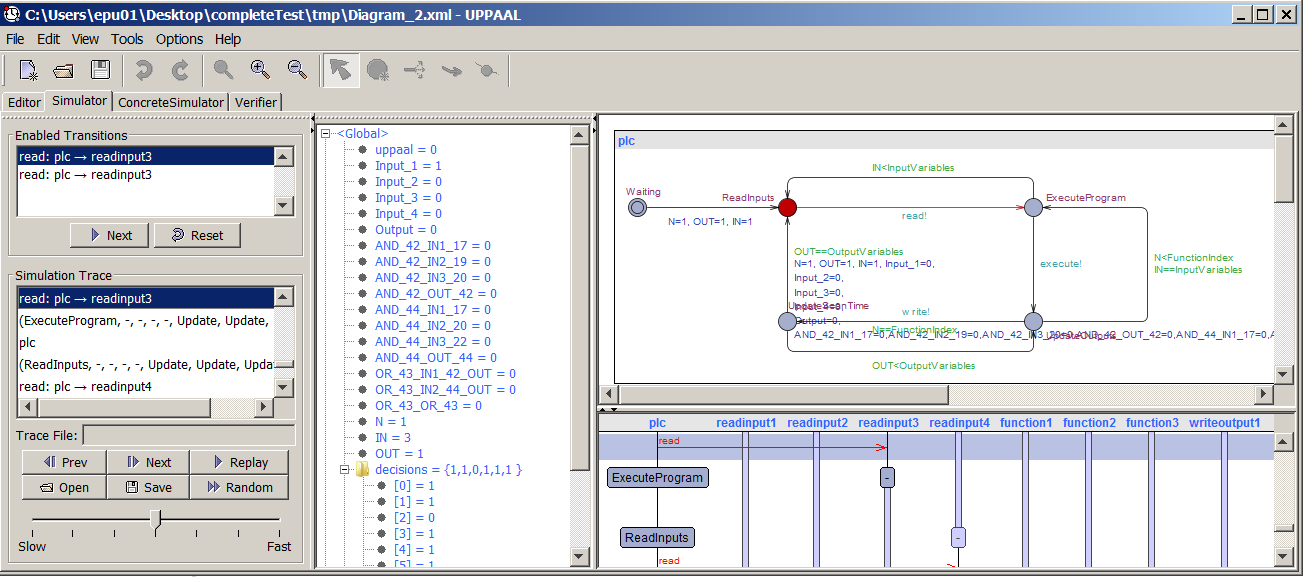}
                \caption{{\uppaal} Interface}
                \label{fig:uppaal}
        \end{subfigure}%
        \\
        \begin{subfigure}[b]{\textwidth}
                \includegraphics[width=\textwidth]{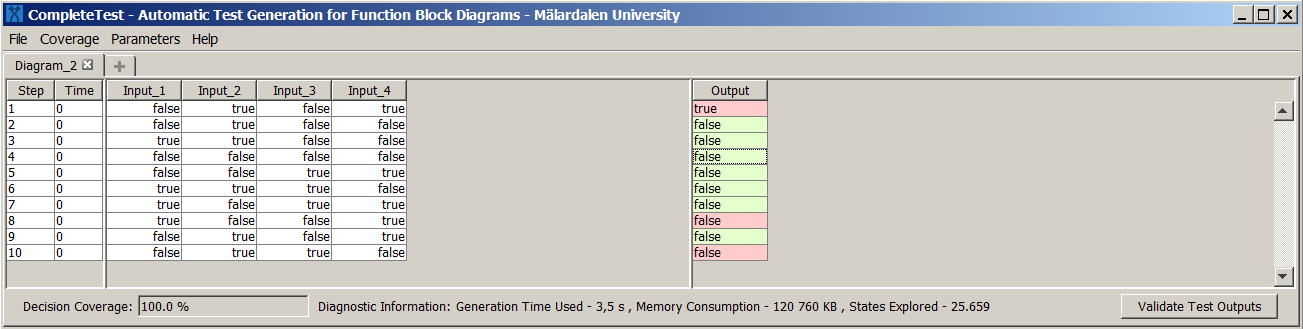}
                \caption{{\completeTest} Interface}
                \label{fig:completetest}
        \end{subfigure}
        \caption{View of the {\uppaal} academic Tool and the {\completeTest} Tool}\label{fig:tool}
\end{figure}

We observed that open sharing of information for research purposes is not only a vital part of the internal decisions, but it is dependent on the continuous interaction with the academic partner. We show in Table \ref{tab:open} a list of materials we had access to and could be used in our studies.
\begin{table}
\renewcommand{\arraystretch}{1.0}
\caption{Open Shared Information Accessed during our Industrial Collaboration.}
\label{tab:open}
\centering
\begin{tabular}{p{0.9\linewidth}}
    \hline
    \\
Material about the testing process.  \\
Software artifacts (e.g., design material, code).\\
Testing artifacts (e.g., Test specifications, requirements, test cases). \\
Development and testing tools.\\
Results of the experiments.\\
    \\
    \hline
\end{tabular}
\end{table}
The engineers we collaborated with were aware of the shared scientific culture, that is to exchange material for research purposes. This was not impeded by the managers acting in the project, and was in fact encouraged by them. This indicates that the company showed a greater openness by drawing results directly from the scientific studies. As shown in Figure \ref{fig:timeline} between 2013 and 2014 we successfully conducted several studies published in peer-reviewed workshops and conferences. Of course a company cannot show naive openness, but we observed during the two year relationship that the company is adapting to academic disclosure practices in exchange for valuable knowledge.
    \\  \\
\subsection{Tool Support}
We observed that from the beginning of the project in 2012, the tool support was necessary for the full industrial usage. Three issues were encountered while developing the tool which we discovered can greatly enhance the success story: flexibility, tool licensing and advanced tool capabilities.

Firstly, the purpose of the tool support is to be useful to the academic studies and in industrial adoption. Flexibility is necessary such that new features can be added without the fear of interfering with current industrial needs. Secondly, the licensing is another issue that influenced how the engineers continuously collaborated with the university. {\completeTest} is an academic tool solely used for the scope of obtaining academic research results. The company was aware of this license and decided to intensively collaborate with the university such that the results of our studies could be of benefit to the industrial practice. Thirdly, the tool was implemented on top of an academic model-checker (see Figure \ref{fig:uppaal}) such that test generation worked seamlessly. In Figure \ref{fig:tool} we show both the academic {\uppaal} model checker and the {\completeTest} testing tool. As model-checkers are hard to use directly into industrial application, additional skills are needed to be build within the organization such that an engineer can automatically generate tests from a formal model. We bi-passed this by developing a tool shown in  Figure \ref{fig:completetest} on top of the model-checker such that an engineer would use the tool directly without any previous knowledge of model-checking. 

We observed that engineers were more likely to use the technology when the testing tool was packaged as part of a larger and familiar interface. In this case the model checker served as a carrier of useful academic concepts. We note that tool support was always maintained by sustained guidance during the usage of the tool.

There are problems, however, with evaluating the usage of the tool support. We cannot yet estimate the costs of application of the tool that uses these new automated test generation approach. Using existing practice, one can use measures for costing at various stages of development.

\subsection{Weekly Progress Meetings}

Weekly meetings were held to ensure project progress and to receive feedback with regard to the feasibility of the studies. Separate meetings arranged by the company were set up to discuss industrial integration of research. Effective communication, the availability of industrial requirements and the open access to an industrial system-under-test were fundamental to a successful outcome of the meetings.

One of the major achievements was that automatic test generation technology was designed together with people working in the company during these meetings. For example, the tool was defined to use developed models and to transform them automatically into a formal model to be used by the model checker. The process took into account the verification that would have to be performed, as well as eventual unit testing. It is important to note that by having weekly collaboration we served practical purposes that would enable engineers to test new ideas almost immediately. 

\subsection{Internal Promotion}
Effective communication and promotion of the benefits of using the {\completeTest} approach to engineers knowledgeable in software testing was an important enabling factor for collaboration. The combination of graphical notations and clear test results that were close to the engineering notations used by engineers provided an effective internal promotion platform to start with. 

The importance of promoting {\completeTest}'s ease of use cannot be underestimated in developing a successful automated test generation technique. Testing safety-critical software is already a complex activity; we do not need to augment the problem with unnecessary complex technologies.

\section{Impediments}
During the study, it was clear that some factors were perceived as the source of impediments for the research collaboration. As shown in Table \ref{tab:imp} not assuming stable research focus for conducting relevant experiments, lack of knowledge of techniques and tools evaluated already in academia, and resistance to change because of opinions and anecdotal evidence are the perceived impediments during our long-term research collaboration.

One of the impediments to collaboration research during the case study has been the difficulty in communicating with engineers that are not comfortable with academic techniques and tools. Perhaps, we need to distinguish between the tools (and underlying techniques) that are used amongst academic experts and those that we use to communicate with engineers who are not expert in automated test generation. This issue appeared in our case study. Substantial effort was required to present the technique in a form understandable to the uninitiated. 

The majority of the observed impediments were related to work performed in 2013 and 2014. For example not assuming a stable research focus was related to work new to the company engineers.  This is not a surprising phenomenon, if one considers that in production engineers and testers are starting quite often new features, while the research is concentrating on the previously started features. This means that we faced the choice of either concentrating on new goals and risk whatever work is ongoing on the current studies, or keep to research commitments.

We observed that due to anecdotal evidence and personal communication the research focus was influenced multiple times. There is a clear need for improved integration of academic evidence with the industrial practice. Successful integration of academic evidence is important to the long-term success of collaboration research. An industrial partner will probably not abandon its current practices, but it is willing to change and enhance its practices.

\begin{table}
\renewcommand{\arraystretch}{1.0}
\caption{Perceived Impediments during our collaborative effort.}
\label{tab:imp}
\centering
\begin{tabular}{p{0.01\linewidth} p{0.85\linewidth}}
    \hline
    \\
1 & Lack of knowledge of techniques and tools evaluated in academia.    \\
2 & Not assuming stable research focus for the conduct of relevant experiments.  \\
3 & Resistance to change: Opinions and anecdotal evidence influencing the practice.  \\
    \hline
\end{tabular}
\end{table}

\section{Conclusion and Future Work}
Collaborative research is an effective tool to promote scientific results in software testing if appropriately implemented. For two years we developed an automated test generation approach collaboratively with Bombardier Transportation AB. This paper examines enablers and impediments for an efficient and effective industry-academia collaboration in the software testing domain. We found that technology is a strong enabler to better collaboration. In addition other enabling factors are recognized like industrial openness to disclose research results, the use of a dedicated tooling platform, and a stable research focus. 

In order to gather more knowledge on industry-academia collaboration enablers we plan to perform a survey with engineers, testers, and managers in the company such that we have a better understanding of the success factors we identified in this study. We propose that further research into the benefits and disadvantages of industrial openness to disclose research results should be performed.

\section{Acknowledgments}
This research was supported by VINNOVA, the Swedish Governmental Agency for Innovation Systems within the ATAC project and The Knowledge Foundation (KKS) through the ITS-EASY industrial research school.

\bibliographystyle{plain}
\bibliography{sigproc}

\begin{thebibliography}{10}

\bibitem{ali2010systematic}
Shaukat Ali, Lionel~C Briand, Hadi Hemmati, and Rajwinder~Kaur
  Panesar-Walawege.
\newblock A systematic review of the application and empirical investigation of
  search-based test case generation.
\newblock {\em IEEE Transactions on Software Engineering}, 36(6):742--762,
  2010.

\bibitem{atacsota}
ATAC~Consortium Contributors.
\newblock Processes and methods for advanced test automation of complex
  software-intensive systems.
\newblock {\em ITEA 2 10037}, 1(3), 2014.

\bibitem{czarnitzki2014delay}
Dirk Czarnitzki, Christoph Grimpe, and Andrew~A Toole.
\newblock Delay and secrecy: Does industry sponsorship jeopardize disclosure of
  academic research?
\newblock {\em Industrial and Corporate Change}, 2014.

\bibitem{dooley2007university}
Lawrence Dooley and David Kirk.
\newblock University-industry collaboration: grafting the entrepreneurial
  paradigm onto academic structures.
\newblock {\em European Journal of Innovation Management}, 10(3):316--332,
  2007.

\bibitem{eldh2013some}
Sigrid Eldh.
\newblock Some researcher considerations when conducting empirical studies in
  industry.
\newblock In {\em 2013 1st International Workshop on Conducting Empirical
  Studies in Industry (CESI)}, pages 69--70. IEEE, 2013.

\bibitem{enoiu2013mos}
Eduard~Paul Enoiu, Kivanc Doganay, Markus Bohlin, Daniel Sundmark, and Paul
  Pettersson.
\newblock Mos: An integrated model-based and search-based testing tool for
  function block diagrams.
\newblock In {\em 2013 1st International Workshop on Combining Modelling and
  Search-Based Software Engineering (CMSBSE)}, pages 55--60. IEEE, 2013.

\bibitem{enoiu2013model}
Eduard~Paul Enoiu, Daniel Sundmark, and Paul Pettersson.
\newblock Model-based test suite generation for function block diagrams using
  the uppaal model checker.
\newblock In {\em 2013 IEEE Sixth International Conference on Software Testing,
  Verification and Validation Workshops (ICSTW)}, pages 158--167. IEEE, 2013.

\bibitem{enoiu2013using}
Eduard~Paul Enoiu, Daniel Sundmark, and Paul Pettersson.
\newblock Using logic coverage to improve testing function block diagrams.
\newblock In {\em Testing Software and Systems}, pages 1--16. Springer, 2013.

\bibitem{glass1994assessment}
Robert~L Glass.
\newblock An assessment of systems and software engineering scholars and
  institutions.
\newblock {\em Journal of Systems and Software}, 27(1):63--67, 1994.

\bibitem{lee1996technology}
Yong~S Lee.
\newblock Technology transfer and the research university: a search for the
  boundaries of university-industry collaboration.
\newblock {\em Research policy}, 25(6):843--863, 1996.

\bibitem{lee2000sustainability}
Yong~S Lee.
\newblock The sustainability of university-industry research collaboration: an
  empirical assessment.
\newblock {\em The Journal of Technology Transfer}, 25(2):111--133, 2000.

\bibitem{perkmann2007university}
Markus Perkmann and Kathryn Walsh.
\newblock University--industry relationships and open innovation: Towards a
  research agenda.
\newblock {\em International Journal of Management Reviews}, 9(4):259--280,
  2007.

\bibitem{perkmann2008engaging}
Markus Perkmann and Kathryn Walsh.
\newblock Engaging the scholar: Three types of academic consulting and their
  impact on universities and industry.
\newblock {\em Research Policy}, 37(10):1884--1891, 2008.

\end{thebibliography}
\end{document}